\documentclass[prd,longbibliography,twocolumn,amsmath,amssymb,nofootinbib,preprintnumbers,balancelastpage]{revtex4-1}

\usepackage{feynmp-auto}
\usepackage{amsmath}
\usepackage[utf8x]{inputenc} 

\usepackage{bbm}
\usepackage{soul}	
\usepackage{xcolor}

\usepackage{epsfig}
\usepackage{color}
\usepackage{slashed}
\usepackage{comment}
\graphicspath{ {Figures/} }
\usepackage[font={small}]{caption}
\usepackage{float}
\usepackage{multirow}
\usepackage[export]{adjustbox}
\usepackage[hang, flushmargin,bottom]{footmisc} 
\usepackage{hyperref}
\hypersetup{
    pdfnewwindow=true,      
    colorlinks=true,        
    linkcolor=blue,        
    citecolor=blue,         
    filecolor=blue,         
    urlcolor=blue           
}

\usepackage{placeins}
\usepackage{enumitem}

\newcommand{\beq}{\begin{equation}}
\newcommand{\eeq}{\end{equation}}

\newcommand{\SU}{\,{\rm SU}}
\newcommand{\U}{\,{\rm U}}
\newcommand{\SO}{\,{\rm SO}}

\usepackage[normalem]{ulem}

\DeclareUnicodeCharacter{2212}{-}

\AtBeginDocument{%
    \newwrite\bibnotes
    \def\bibnotesext{Notes.bib}
    \immediate\openout\bibnotes=\jobname\bibnotesext
    \immediate\write\bibnotes{@CONTROL{REVTEX41Control}}
    \immediate\write\bibnotes{@CONTROL{%
    apsrev41Control,author="08",editor="1",pages="1",title="0",year="1"}}
     \if@filesw
     \immediate\write\@auxout{\string\citation{apsrev41Control}}%
    \fi
}%

\usepackage{amsfonts}
\usepackage{graphicx}
\usepackage{tcolorbox}
\usepackage{amsmath}
\usepackage{slashed}
\usepackage{verbatim}
\usepackage{color}
\usepackage{array}
\newcolumntype{P}[1]{>{\centering\arraybackslash}p{#1}}
\usepackage{physics}
\usepackage[font={small, up}]{caption}
\usepackage{hyperref}	

\begin{document}

\title{ {\bf{Probing Quark-Lepton Unification with Leptoquark and Higgs Decays}}}
\author{Pavel Fileviez P\'erez}
\email{pxf112@case.edu}
\affiliation{Physics Department and Center for Education and Research in Cosmology and Astrophysics (CERCA), 
Case Western Reserve University, Cleveland, OH 44106, USA}
\author{Elliot Golias}
\email{elliot.golias@case.edu}
\affiliation{Physics Department and Center for Education and Research in Cosmology and Astrophysics (CERCA), 
Case Western Reserve University, Cleveland, OH 44106, USA}
\author{Alexis D. Plascencia}
\email{alexis.plascencia@case.edu}
\affiliation{Physics Department and Center for Education and Research in Cosmology and Astrophysics (CERCA), Case Western Reserve University, Cleveland, OH 44106, USA}
\begin{abstract}
We point out unique relations between the decay widths for leptoquarks and Higgs bosons that can be used to test the unification of quarks and leptons at the TeV scale. We discuss the main predictions of the minimal theory for quark-lepton unification and show how the different decays for leptoquarks and Higgses are related by the symmetry of the theory.
\end{abstract}
\maketitle
\hypersetup{linkcolor=blue}
%
\section{Introduction}
After the discovery of the Standard Model (SM) Higgs boson at the Large Hadron Collider (LHC) we 
know that the SM of Particle Physics can describe with high precision the physics below the TeV scale.
Nonetheless, there are reasons to believe that the LHC could discover new forces and 
a new sector that could help address some of the open issues in particle physics and cosmology. 

It is well-known that the SM needs to be modified in order to explain the origin of neutrino masses, the nature of dark matter and the baryon asymmetry in the Universe. Unfortunately, we do not know what is the cut-off scale of the SM and there is no certainty that the LHC will discover new physics. The unification of forces in nature is one 
of the best ideas we have for physics beyond the Standard Model. The simplest unified theories 
based on $\SU(5)$ and $\SO(10)$ can describe physics at the high scale, $M_{\rm GUT} \sim 10^{15-16}$ GeV, for reviews see Refs.~\cite{Senjanovic:2006zz, Nath:2006ut}, and we cannot hope to directly test their predictions at colliders because the fields are superheavy.

J. Pati and A. Salam~\cite{Pati:1974yy} proposed the idea of matter unification, where the SM quarks and leptons can be unified in the same multiplet. 
This idea was crucial to understand the idea of grand unification.
The minimal Pati-Salam model predicts the same mass for neutrinos and up-quarks, and the same for down-quarks and charged leptons.
Therefore, if we use the canonical seesaw 
mechanism~\cite{Minkowski:1977sc,Yanagida:1979as,GellMann:1980vs,Mohapatra:1979ia} the relevant scale is $M_R \sim 10^{14}$ GeV.

In our view, the generic idea of quark-lepton unification is very appealing. Some years ago, a simple theory 
for the unification of quarks and leptons below the multi-TeV scale was proposed by one of the authors, P. Fileviez Perez, and M. B. Wise in Ref.~\cite{FileviezPerez:2013zmv}.  This theory is based 
on $\SU(4)_C \otimes \SU(2)_L \otimes \U(1)_R$, and neutrino masses are generated through the inverse 
seesaw mechanism~\cite{Mohapatra:1986aw,Mohapatra:1986bd} in order to have a consistent theory where $\SU(4)_C$ is broken at the low scale.
This theory tells us that one can hope to test the idea of quark-lepton unification at current or future colliders.

In this article, motivated by the possibility to test the idea of matter unification at colliders, we investigate in detail the Higgs and leptoquark decays in the theory proposed 
in Ref.~\cite{FileviezPerez:2013zmv}. This theory predicts three leptoquarks and two Higgs doublets. The predicted leptoquarks are 
$X_\mu \sim (\mathbf{3},\mathbf{1},2/3)_\text{SM}$, $\Phi_3 \sim (\mathbf{\bar{3}}, \mathbf{2},-1/6)_\text{SM}$ and  $\Phi_4 \sim (\mathbf{3}, \mathbf{2}, 7/6)_{\rm SM}$. We point out relations between the decay widths for Leptoquarks and Higgs bosons that can be used to test the idea of quark-lepton unification at the TeV scale.  

In the case of the scalar leptoquarks we find several unique relations for the decay widths, for example:
\begin{eqnarray}
\sum_{i,j=1}^3  \Gamma (\phi_3^{1/3} \to \bar{d}_i \nu_j) &=& \left( \frac{M_{\phi_3^{1/3}} }{M_{\phi_4^{2/3}}} \right) \sum_{i,j=1}^3   \Gamma (\phi_4^{2/3} \to \bar{e}_i d_j). \nonumber
\end{eqnarray}
Therefore, if these decays are measured it can be checked whether this relation predicted by quark-lepton unification holds.
In the above equation the leptoquarks $\phi_3^{1/3} $ and $\phi_4^{2/3}$ are components of the $\Phi_3$ and $\Phi_4$ fields, respectively.

The Higgs sector of this theory is special because there are two Higgs doublets with only four different Yukawa couplings.
One finds, for example, that only two Yukawa couplings determine the masses for the down-quarks and charged leptons. 
In the limit where there is a hierarchy between the two Higgs vacuum expectation values, $\tan \beta = v_2/v_1 \gg 1$, we find that 
the quark-lepton unification predicts the following relation for the heavy CP-even Higgs decay widths
\beq
\sum_{i,j=1}^3 \Gamma( H \to \bar{d}_i d_j) = 3 \sum_{i,j=1}^3 \Gamma( H \to \bar{e}_i e_j).  \nonumber
\eeq
This relation is quite unique because naively we expect the Higgs decays into leptons and quarks to be different. 
Finally, we study the relations between the leptoquark and Higgs decays when $\tan \beta \ll 1$. For example, we find:
\begin{eqnarray}
\sum_{i,j=1}^3 \Gamma (\phi_3^{-2/3} \to \bar{d}_i e_j)  = 4 \left( \frac{M_{\phi_3^{-2/3}}} {M_H}\right) \sum_{i,j=1}^3 \Gamma (H \to \bar{e}_i e_j). \nonumber 
\end{eqnarray}
We discuss how the above relations and other relations can be used to test the idea of quark-lepton unification at colliders. 

This article is organized as follows: in Section~\ref{sec:theory} we discuss the minimal theory for quark-lepton unification at the low scale.
In Section~\ref{sec:decays} we discuss the properties of the different leptoquarks and their decays. In Section~\ref{sec:higgsdecays} the new Higgses are discussed and we point out the main properties of their decays into fermions. Our main findings are summarized in Section~\ref{sec:summary}.
\section{Minimal Quark-Lepton Unification}
\label{sec:theory}
A simple renomalizable theory for quark-lepton unification was proposed in Ref.~\cite{FileviezPerez:2013zmv} which can be seen as a low energy limit of the Pati-Salam theory.
This theory is based on the gauge symmetry, $\SU(4)_C \otimes \SU(2)_L \otimes \U(1)_R$ and the SM matter fields are unified in three representations:
\begin{eqnarray}
F_{QL} &=&
\left(
\begin{array}{cc}
u 
&
\nu 
\\
d &  
e
\end{array}
\right) \sim (\mathbf{4}, \mathbf{2}, 0), 
\\
F_u &=&
\left(
\begin{array}{cc}
u^c   
&
\nu^c
\end{array}
\right) \sim (\mathbf{\bar{4}}, \mathbf{1}, -1/2), 
\\
F_d &=&
\left(
\begin{array}{cccc}
d^c  
&
e^c
\end{array}
\right) \sim (\mathbf{\bar{4}}, \mathbf{1}, 1/2),
\end{eqnarray}
while the gauge fields live in $A_\mu \sim (\mathbf{15}, \mathbf{1},0)$. The minimal Higgs sector has three scalar representations: $\Phi \sim (\mathbf{15}, \mathbf{2}, 1/2)$, $\chi \sim  (\mathbf{4}, \mathbf{1}, 1/2)$ and $H_1 \sim  (\mathbf{1}, \mathbf{2}, 1/2)$. This minimal sector allows us to write a full realistic theory for matter unification.

The gauge symmetry, $\SU(4)_C \otimes \SU(2)_L \otimes \U(1)_R$,  is spontaneously broken to the SM gauge group by the vacuum expectation value (VEV) of the scalar
field $\chi$, i.e. $\langle \chi \rangle = \textrm{diag} ( 0, \, 0, \, 0, \, v_{\chi}/\sqrt{2})$, which gives mass to the vector leptoquark $X_\mu$, defining the scale of matter unification. See Appendix~\ref{app:fields} 
for more details. 

The Yukawa interactions for the charged fermions can be written as
\begin{eqnarray}
- {\cal L} &\supset&
Y_1  \, {F}_{QL} F_u H_1  \ + \ Y_2 \,  {F}_{QL} F_u \Phi  \nonumber \\
&+ &  Y_3 \,  H_1^\dagger {F}_{QL} F_d  \ + \  Y_4  \, \Phi^\dagger {F}_{QL}  F_d   + \mbox{h.c.},
\label{eq:Yukawa}
\end{eqnarray}
while for neutrinos one can implement the inverse seesaw mechanism using the terms
\begin{eqnarray}
- {\cal L} &\supset&
Y_5 F_u \chi S  \ + \  \frac{1}{2} \mu S S   + \mbox{h.c.}.
\end{eqnarray} 
Here the fields $S \sim (\mathbf{1}, \mathbf{1}, 0)$ are SM fermionic singlets.
In order to achieve very small neutrino masses one needs a seesaw mechanism.
The minimal scenario to have the $\SU(4)_C$ symmetry broken at the low scale and generate small neutrino masses without fine-tuning 
is using the inverse seesaw mechanism~\cite{FileviezPerez:2013zmv}. This is a key idea that allows us to realize matter unification below the multi-TeV scale.

The mass matrix for neutrinos in the basis ($\nu$, $\nu^c$, $S$) reads as
\begin{equation}
\left( \nu \  \nu^c \  S  \right) 
\left(\begin{array}{ccc} 
0 & M_\nu^D & 0  \\ 
(M_\nu^D)^T & 0 & M_\chi^D \\
0 &  (M_\chi^D)^T & \mu
\end{array}\right)  
\left(\begin{array}{c} \nu \\  \nu^c \\ S  \end{array}\right),
\end{equation}
where
$M_\chi^D = Y_5 \, v_\chi / \sqrt{2}$ and $M_\nu^D$ is given in Eq.~(\ref{fermionmasses}).

The light neutrino masses are given by 
\begin{equation}
m_\nu \approx \mu \, (M_\nu^D)^2 / (M_\chi^D)^2,
\end{equation}
when $M_\chi^D \gg M_\nu^D \gg \mu$ and the heavy neutrinos form a pseudo-Dirac pair.

In order to test the generic idea of quark-lepton unification we need to understand the predictions of the different theories where this idea is realized.
In this article, we focus on the minimal theory for matter unification that can describe physics below the multi-TeV scale, because we can 
hope to test this idea at current or future colliders. For phenomenological studies of this theory see Refs.~\cite{Smirnov:1995jq,Faber:2018qon,Faber:2018afz,Perez:2021ddi}.
\section{Leptoquark Decays}
\label{sec:decays}
The theory discussed in the previous section predicts a vector leptoquark, $X_\mu \sim (\mathbf{3},\mathbf{1},2/3)_\text{SM}$, associated to the $\SU(4)_C$ symmetry, and four physical scalar leptoquarks.
The scalar leptoquarks $\Phi_3 \sim (\mathbf{\bar{3}}, \mathbf{2},-1/6)_\text{SM}$ and $\Phi_4 \sim (\mathbf{3}, \mathbf{2}, 7/6)_{\rm SM}$ can be written in $\SU(2)_L$ components as,
\begin{equation}
\Phi_3 =  \begin{pmatrix} \phi_3^{1/3} \\[1ex]\phi_3^{-2/3} \end{pmatrix},
 \hspace{0.5cm} \text{and} \hspace{0.5cm} \Phi_4 =\begin{pmatrix} \phi_4^{5/3} \\[1ex]\phi_4^{2/3} \end{pmatrix},
\end{equation}
where the numbers in the superscript denote the electric charge. The Yukawa interactions for $\Phi_3$ and $\Phi_4$ are given by
\begin{eqnarray}
\label{eq:YukInt}
-\mathcal{L} &\supset& Y_2 \, \varepsilon_{ab} \, \ell_L^a \, \Phi_4^b \, (u^c)_L + Y_2 \, \varepsilon_{ab} \, Q_L^a \, \Phi_3^b \, (\nu^c)_L \nonumber \\
&+ & Y_4 \, \Phi_3^\dagger \, \ell_L \, (d^c)_L + Y_4  \, \Phi_4^\dagger  \, Q_L \, (e^c)_L     + \mbox{h.c.} \, ,
\end{eqnarray}
where $a$ and $b$ correspond to the $\SU(2)_L$ indices. Notice that in this sector we only have two different Yukawa couplings because the $\SU(4)_C$ symmetry relates the different Yukawa interactions in a unique way.

The physical scalar leptoquarks in this theory are:
$$\phi_3^{1/3}, \,\,\, \phi_4^{5/3}, \,\,\, \phi_3^{-2/3} \,\,\, \textrm{and} \,\,\, \phi_4^{2/3}.$$
The leptoquarks $\phi_3^{-2/3}$ and $\phi_4^{2/3}$ can mix but the mixing angle is determined by the electroweak scale, and hence, it is generically very small. Consequently, in this work we ignore this mixing.
For the interactions of these fields see Appendix~\ref{leptoquarks-rules}. 

Now, let us discuss the different decays of the leptoquarks and their decays to understand how the quark-lepton symmetry predicts unique relations 
between the different decay widths.

{\boldmath$X_\mu$} \textbf{\textit{decays}}: The vector leptoquark $X_\mu$ can have the following decays
\begin{equation} 
X_\mu \to \bar{e}_i d_j, \,\, \bar{\nu}_i u_j, \nonumber
\end{equation} 
where $i,j=1,2,3$ correspond to the family indices. Unfortunately, naively one expects that the 
vector leptoquark mass must be above $10^3$ TeV to satisfy 
the experimental bounds on rare decays such as $K_L \to e^\pm \mu^{\mp}$~\cite{Valencia:1994cj}, 
unless one uses the freedom on the mixings between quarks and leptons.
For the vector leptoquark there is a very simple relation for the decay widths:
\begin{eqnarray}
\sum_{i.j=1}^3 \Gamma (X_\mu \to \bar{e}_i d_j) &=& 2 \sum_{i,j=1}^3 \Gamma (X_\mu \to \bar{\nu}_i u_j).
\end{eqnarray}
Notice that the total widths for a given decay channel are clean, meaning that they are independent of the unknown mixing angles between quarks and leptons determined by the matrix $V_{DE}$ defined in Appendix~\ref{leptoquarks-rules}. If the right-handed neutrinos are much lighter than the $X_\mu$ boson, the decay widths of $X_\mu$ into quarks and leptons are equal.

\textbf{\textit{Scalar Leptoquarks}}: In the case of the scalar leptoquarks one can have the decays
\begin{eqnarray}
\phi_3^{1/3} &\to& \bar{d}_i \nu_j, \, \bar{d}_i N_j; \hspace{1.2cm}
\phi_4^{5/3} \to \bar{e}_i u_j; \nonumber \\
\phi_3^{-2/3} &\to& \bar{d}_i e_j, \, \bar{u}_i \nu_j, \,  \bar{u}_i N_j; \, \, \,\,\,
\phi_4^{2/3} \to \bar{e}_i d_j, \, \bar{\nu}_i u_j, \,  \bar{N}_i u_j. \nonumber
\end{eqnarray}
Even though the Feynman rules for the scalar leptoquarks listed in Appendix~\ref{leptoquarks-rules} are involved, it is possible to define some total widths for different channels 
that are independent of the unknown mixing angles entering in the interactions. 
For example, by defining:
\begin{align}
\Gamma_T(\phi_3^{1/3} \to \bar{d} \nu ) &\equiv \sum_{i,j=1}^3 \Gamma(\phi_3^{1/3} \to \bar{d}_i \nu_j) \nonumber \\[1ex]
  &= \frac{ 3  M_{\phi_3^{1/3}}}{16 \pi} {\rm Tr} [ Y_4^\dagger Y_4 ] ,
\label{phi3decay}
\end{align}
which turns out to be independent of any mixing angle when the fermion masses are neglected, see Appendix~\ref{app:decaywidths} for the details of the calculation. In order to understand this simple result, notice that the individual decay width $\Gamma(\phi_3^{1/3} \to \bar{d}_i \nu_j)$ 
is a complicated function of mixing matrices and Yukawa couplings. See the Feynman rules in Appendix~\ref{leptoquarks-rules} for the details. However, since the leptoquarks are heavy we can neglect the fermion masses and by summing over the family index we find a total decay width that is a function only of the Yukawa couplings. This simple idea allows us to find relations between the different decays.

We investigated the different leptoquark decays into fermions and found the following set of relations predicted by quark-lepton unification: 
\begin{eqnarray}
\frac{\Gamma_T (\phi_3^{1/3} \to \bar{d} \nu)}{M_{\phi_3^{1/3}}} &=& \frac{\Gamma_T (\phi_4^{2/3} \to \bar{e} d)}{M_{\phi_4^{2/3}}} \nonumber \\
&=& \frac{\Gamma_T (\phi_3^{-2/3} \to \bar{d} e)}{M_{\phi_3^{-2/3}}}.
\end{eqnarray}
Now, if the decay channels with the heavy pseudo-Dirac neutrinos are available, neglecting all fermion masses one finds:
\begin{eqnarray}
\frac{\Gamma_T (\phi_3^{1/3} \to \bar{d} N)}{M_{\phi_3^{1/3}}} &=& \frac{\Gamma_T (\phi_4^{2/3} \to \bar{\nu} u)}{M_{\phi_4^{2/3}}} \nonumber \\
&=& \frac{\Gamma_T (\phi_3^{-2/3} \to \bar{u} N)}{M_{\phi_3^{-2/3}}},\\[1ex]
\Gamma_T (\phi_4^{5/3} \to \bar{e} u ) &=&  \frac{M_{\phi_4^{5/3}}}{M_{\phi_3^{-2/3}}} \left[ \Gamma_T (\phi_3^{-2/3} \to \bar{u} N )\right. \nonumber \\
&&\left. + \, \Gamma_T (\phi_3^{-2/3} \to \bar{d} e ) \right]. 
\end{eqnarray}

Furthermore, if the pseudo-Dirac neutrinos are light and when $\phi_3^{1/3}$ and $\phi_4^{5/3}$ are the lightest elements of the $\Phi_3$ and $\Phi_4$   respectively; then, their total widths will be dominated by the $\ell + q$ decays, and hence, their lifetimes will satisfy
\beq
 M_{\phi_3^{1/3}} \,\, \tau_{\phi_3^{1/3}} = M_{\phi_4^{5/3}} \,\, \tau_{\phi_4^{5/3}} \,\, .
\eeq
Similarly, if $\phi_3^{1/3}$ and $\phi_4^{2/3}$ are the lightest elements of the $\Phi_3$ and $\Phi_4$  we then have
\beq
 M_{\phi_3^{1/3}} \,\, \tau_{\phi_3^{1/3}} = M_{\phi_4^{2/3}} \,\, \tau_{\phi_4^{2/3}} \,\, .
\eeq
The same is true when $\phi_3^{-2/3}$ and either $\phi_4^{2/3}$ or $\phi_4^{5/3}$ are the lightest elements
\begin{align}
 M_{\phi_3^{-2/3}} \,\, \tau_{\phi_3^{-2/3}} = M_{\phi_4^{2/3}} \,\, \tau_{\phi_4^{2/3}} \,\, , \\
  M_{\phi_3^{-2/3}} \,\, \tau_{\phi_3^{-2/3}} = M_{\phi_4^{5/3}} \,\, \tau_{\phi_4^{5/3}} \,\, .
\end{align}
Clearly, these relations are predictions from the unification of quark and leptons and can be used to test this idea at particle colliders.

It is well-known that any theory, including the SM, does not predict the values of the gauge and Yukawa couplings present in the interactions, but can predict relations between the different physical quantities. In this case, the minimal theory for quark-lepton unification is predicting a set of relations for the decay widths that can be tested if these leptoquarks are discovered in the near future.

It is important to mention that the branching ratios for the leptoquark decays depend of the mass splittings of the different components of $\Phi_3$ 
and $\Phi_4$. For example,  $\phi_3^{1/3}$ can have the following decays with the $\phi_3^{-2/3}$ in the final state:  
$\phi_3^{1/3} \to W^+ \phi_3^{-2/3}, \,\, \pi^+ \phi_3^{-2/3}, \,\, \bar{e}_i \nu_j \phi_3^{-2/3}$. Only when the mass splitting is small 
one can a large branching ratio for $\phi_3^{1/3} \to \bar{d}_i \nu_j$. For a study about the relation between the decays of a $\SU(2)_L$ doublet leptoquark
see the studies in Ref.~\cite{FileviezPerez:2008dw}. See also Ref.~\cite{Dorsner:2016wpm} for a review about leptoquarks and Ref.~\cite{Murgui:2021bdy} for a 
recent discussion about proton decay mediated by scalar Leptoquarks in this theory, where the authors have shown that there are no dimension five contributions to proton decay.
\section{Higgs Decays}
\label{sec:higgsdecays}
This theory predicts a unique Higgs sector with two Higgs doublets, $H_1 \sim (\mathbf{1}, \mathbf{2}, 1/2)$ and $H_2 \sim (\mathbf{1}, \mathbf{2}, 1/2)$, 
with {\textit{only four}} independent Yukawa couplings. The Yukawa couplings for the Higgses can 
be written as
\begin{eqnarray}
- \mathcal{L}_Y & = & \bar{u}_R \left( Y_1^T H_1 +  \frac{1}{2 \sqrt{3}}  Y_2^T H_2 \right) Q_L  \nonumber \\
&+&  \bar{N}_R \left( Y_1^T H_1 -  \frac{\sqrt{3}}{2}  Y_2^T H_2 \right)  \ell_L  \nonumber \\ 
&+ & \bar{d}_R \left( Y_3^T H_1^\dagger   + \frac{1}{2 \sqrt{3}}  Y_4^T H_2^\dagger \right) Q_L  \nonumber \\  
&+ & \bar{e}_R  \left(  Y_3^T H_1^\dagger -  \frac{\sqrt{3}}{2} Y_4^T H_2^\dagger  \right) \ell_L + {\rm h.c.} \, .
\end{eqnarray} 
Here we neglect the small mixing between the Higgs doublets and the $\chi$ field. Notice that in the general two Higgs doublet model (commonly referred in the literature as the type-III 2HDM) there are eight different Yukawa couplings, but in our case the $\SU(4)_C$ symmetry relates quarks and leptons so there are only four independent Yukawa couplings. For reviews on two-Higgs doublet models we refer the reader to Refs.~\cite{Gunion:1989we,Branco:2011iw}.

After symmetry breaking the charged fermions and the Dirac neutrino masses are given by
\begin{eqnarray}
M_U &=& Y_1 \frac{ v_1}{\sqrt{2}} + \frac{1}{2 \sqrt{3}} Y_2 \frac{ v_2}{\sqrt{2}}, \\
M_\nu^D &=& Y_1 \frac{ v_1}{\sqrt{2}} - \frac{\sqrt{3}}{2} Y_2 \frac{ v_2}{\sqrt{2}}, 
\label{fermionmasses}
\\
M_D &=& Y_3  \frac{v_1}{\sqrt{2}} + \frac{1}{2 \sqrt{3}} Y_4 \frac{ v_2}{\sqrt{2}}, \\
M_E &=& Y_3\frac{ v_1}{\sqrt{2}} - \frac{\sqrt{3}}{2} Y_4 \frac{ v_2}{\sqrt{2}}.
\end{eqnarray}
Here the VEVs of the Higgs doublets are defined as $\langle H^0_1 \rangle = v_1 / \sqrt{2}$, and $\langle H^0_2  \rangle  = v_2/\sqrt{2}$. 
Notice that the above equations allow us to have different masses for charged leptons and down-quarks. However, 
one needs the inverse seesaw mechanism to generate small neutrino masses discussed in Section~\ref{sec:theory}.

In our convention the mass matrices are diagonalized as
\begin{eqnarray}
&& U^T M_U U_c = M_U^{\rm diag}, \\
&& D^T M_D D_c = M_D^{\rm diag}, \\
&& E^T M_E E_c = M_E^{\rm diag}.
\end{eqnarray}
Notice that the above relations tell us that the Yukawa coupling $Y_4$ defines the difference between $M_E$ and $M_D$, and one 
can use these relations to write the decay widths for leptoquarks as a function of quark masses. For example,
\begin{eqnarray}
\Gamma_T(\phi_4^{2/3} \to \bar{e} d ) &=& \frac{ 3 M_{\phi_4^{2/3}}}{16 \pi} {\rm Tr} [ Y_4^\dagger Y_4 ]  \simeq \frac{ 9  m_b^2 \, M_{\phi_4^{2/3}}}{ 32 \pi \, v_2^2} .
  \label{phi3decay}
\end{eqnarray}
As in any two-Higgs doublet model, assuming CP-conservation,  the physical fields are: $h$ and $H$ the CP-even neutral fields, $A$ the CP-odd field, and 
two charged Higgs bosons $H^{\pm}$. The $h$ field is identified as the SM-like Higgs boson. We list all the Feynman rules in Appendix~\ref{sec:appHiggs}. As in the case of the leptoquarks, one can find a set of relations between the decay widths of the Higgs bosons. 

The physical Higgs fields are defined as
\begin{eqnarray}
\begin{pmatrix} H \\ h \end{pmatrix} = \begin{pmatrix} \cos \alpha & \sin \alpha \\ -\sin \alpha & \cos \alpha \end{pmatrix}
\begin{pmatrix} H_1^0 \\ H_2^0 \end{pmatrix}, \\
\begin{pmatrix} G^0 \\ A^0 \end{pmatrix} = \begin{pmatrix} \cos \beta & \sin \beta \\ -\sin \beta & \cos \beta \end{pmatrix}
\begin{pmatrix} A_1^0 \\ A_2^0 \end{pmatrix}, \\
\begin{pmatrix} G^\pm \\ H^\pm \end{pmatrix} = \begin{pmatrix} \cos \beta & \sin \beta \\ -\sin \beta & \cos \beta \end{pmatrix}
\begin{pmatrix} H_1^\pm \\ H_2^\pm \end{pmatrix},
\end{eqnarray}
where $H^0_i$, $H_i^{\pm}$, $A_i^0$ are the neutral, charged, and CP-odd components of the Higgs doublets, respectively, and $G^0$, $G^{\pm}$ are the Goldstone bosons. Furthermore, the mixing angle $\beta$ is related to the VEVs of the Higgs doublets by $\tan{\beta} = v_2/v_1$.
Since the $h$ field is the SM-like Higgs boson, in order to agree with the measured properties of the SM Higgs field we will work in the limit $\sin (\beta - \alpha) \to 1$ (or $\alpha \simeq \beta - \pi/2$).
In this limit we make sure the couplings of $h$ are SM-like. 

The physical Higgs can have the usual decays to matter:
\begin{align}
H, \, A & \to \bar{u}_i u_j, \,\, \bar{d}_i d_j, \,\, \bar{e}_i e_j, \,\, \bar{\nu}_i N_j;  \nonumber \\ 
H^{+} & \to  \bar{e}_i \nu_j, \,\, \bar{e}_i N_j, \,\, \bar{d}_i u_j. \nonumber
\end{align}
The decay widths of these Higgses are a function of the Yukawa couplings and unknown mixing angles. 

Following the same idea used for the leptoquark decays, we can define the total widths for the different channels and then find simple relations between the Higgs decays. In the limit of $\tan \beta \gg 1$, quark-lepton unification predicts the following unique relation for the heavy Higgs
\beq
\sum_{i,j=1}^3 \Gamma( H \to \bar{d}_i d_j) = 3 \sum_{i,j=1}^3 \Gamma( H \to \bar{e}_i e_j).
\label{Hdecay}
\eeq
Notice that this is a striking relation for the decays into down-quarks and charged leptons.

We can find the following relation for the decay widths of the charged Higgs, $H^{\pm}$, 
if the mass of the right-handed neutrinos, $M_{N_i}$, is much smaller than $M_{H^\pm}$:
\beq
\Gamma_T ( H^+ \to \bar{d} u) = 3 \left[  \Gamma_T ( H^+ \to \bar{e} \nu) + \Gamma_T ( H^+ \to \bar{e} N) \right].
\eeq
A similar relation can be obtained for the CP-odd neutral Higgs:
\beq
\sum_{i,j=1}^3 \Gamma( A \to \bar{d}_i d_j) = 3 \sum_{i,j=1}^3 \Gamma( A \to \bar{e}_i e_j).
\label{Adecay}
\eeq
In the $\tan \beta \ll 1$ limit the relations between the Higgs decays become
\begin{eqnarray}
\Gamma_T ( H \to \bar{e} e) &=&  3 \ \Gamma_T ( H \to \bar{d} d), \label{decaysB}\\[1ex]
\Gamma_T ( A \to \bar{e} e)  &=&  3 \ \Gamma_T ( A \to \bar{d} d), 
\label{decays2}
\end{eqnarray}
and in this limit it is possible to relate the decay widths of leptoquarks with the ones from the Higgs scalars as follows
\begin{align}
\Gamma_T (\phi_4^{2/3} \to \bar{\nu} u) & = \frac{12 \, M_{\phi_4^{2/3}}}{M_H} \, \Gamma_T(H \to \bar{u} u), \\[1ex]
\Gamma_T (\phi_3^{-2/3} \to \bar{d} e) & = 4 \left( \frac{M_{\phi_3^{-2/3}}} {M_H}\right) \Gamma_T (H \to \bar{e} e).
\label{relation1}
\end{align}
The following relation holds independent of the value of $\tan \beta$
\beq
\Gamma_T ( A \to \bar{e} e) = \frac{M_A}{M_H} \Gamma_T ( H \to \bar{e} e). 
\label{relation2}
\eeq
From our perspective, these relations are very unique. Notice that the relations in Eqs.~(\ref{Hdecay})~-~(\ref{relation2})
tell us something interesting: {\textit{the theory predicts simple relations for the total decay widths into quarks and leptons}}. Clearly, these are predictions from the unification of quarks and leptons. We are not aware of any model for physics beyond the SM that can predict these relations for the Higgs and leptoquark decays.

\newpage
\section{Summary}
\label{sec:summary}
The idea of quark-lepton unification is one of the best motivated ideas for physics beyond the Standard Model. We have discussed the minimal gauge theory for quark-lepton unification that can describe physics below the multi-TeV scale. This theory predicts one vector leptoquark, four scalar leptoquarks and a unique Higgs sector with only four independent Yukawa couplings.

In this article we pointed out unique relations between the decay widths for leptoquarks that can 
be used to test the idea of quark-lepton unification. We also discussed the Higgs sector of the theory. The theory has two Higgs doublets with only four different Yukawa couplings determined by the symmetry between quarks and leptons. We discussed the different Higgs decays and showed how the total 
Higgs decay widths into quarks and leptons are related at large (and small) values for the ratio between the vacuum expectation values.
We believe that these results should motivate new studies to test the idea of matter unification at the LHC or future colliders.
\\

{\textit{Acknowledgments:}}
{\small{The work of P.F.P. has been supported by the U.S. Department of Energy, Office
of Science, Office of High Energy Physics, under Award Number DE-SC0020443.}}

\newpage
\appendix
\begin{widetext}
\section{Gauge and Higgs Fields}
\label{app:fields}
%
In this theory the SM gluon fields, the vector leptoquark and the new neutral gauge boson live in the adjoint representation of $\SU(4)_C$:
\begin{eqnarray}
A_\mu =
\left(
\begin{array} {cc}
G_\mu & X_\mu/\sqrt{2}  \\
X_\mu^*/\sqrt{2} & 0  \\
\end{array}
\right) + T_4 \ B_\mu^{'} \sim (\mathbf{15}, \mathbf{1},0),
\end{eqnarray}
where $G_\mu \sim (\mathbf{8},\mathbf{1},0)_\text{SM}$ are the gluons, $X_\mu \sim (\mathbf{3},\mathbf{1},2/3)_\text{SM}$ are vector leptoquarks, and $B_\mu^{'} \sim (\mathbf{1},\mathbf{1},0)_\text{SM}$. 
The Higgs sector is composed of 
\begin{eqnarray}
H^T_1 & = & \left( H^+_1   \  H^0_1 \right) \sim (\mathbf{1}, \mathbf{2}, 1/2), \, \,\,\,\,\,
\chi = \left(  \chi_u  \  \chi_R^0 \right) \sim (\mathbf{4}, \mathbf{1}, 1/2), \, \,\,\,\,\, {\rm{and}} \nonumber \\[1ex]
\Phi &=& 
\left(
\begin{array} {cc}
\Phi_8 & \Phi_3  \\
\Phi_4 & 0  \\
\end{array}
\right) + \sqrt{2} \, T_4 \ H_2 \sim (\mathbf{15}, \mathbf{2}, 1/2).
\end{eqnarray}
Here $H_2 \sim (\mathbf{1}, \mathbf{2}, 1/2)_\text{SM}$ is a second Higgs doublet, $\Phi_8 \sim (\mathbf{8}, \mathbf{2}, 1/2)_\text{SM}$, and the scalar leptoquarks  
$\Phi_3 \sim (\mathbf{\bar{3}}, \mathbf{2},-1/6)_\text{SM}$ and  $\Phi_4 \sim (\mathbf{3}, \mathbf{2}, 7/6)_{\rm SM}$. The $T_4$ generator of $\SU(4)_C$ in the above equation is normalized as
$
T_4 =
\frac{1}{2 \sqrt{6}} \rm{diag} (1,1,1,-3).
$
%
%
\section{Decay Widths}
\label{app:decaywidths}

Here we present the details on how the expressions for the decay widths simplify when we sum over family indices. Let's consider the decay $\phi_3^{1/3}\to \bar{d}_i \nu_j$ and sum over the family indices,
\begin{align}
\Gamma_T(\phi_3^{1/3} \to \bar{d} \nu ) &\equiv \sum_{i,j=1}^3 \Gamma(\phi_3^{1/3} \to \bar{d}_i \nu_j) = \frac{3 M_{\phi_3^{1/3}}}{16 \pi}  (N^* Y_4^\dagger D_c^\dagger)^{ij} (N Y_4^T D_c^T)^{ij} \nonumber \\[1ex]
&= \frac{3 M_{\phi_3^{1/3}}}{16 \pi} N^{\dagger \alpha i } N^{i \sigma} D_c^{\dagger \beta j} D_c^{ j \rho} Y_4^{\dagger \alpha \beta } Y_4^{ \rho \sigma} = \frac{ 3 M_{\phi_3^{1/3}}}{16 \pi} \delta^{\alpha \sigma} \delta^{\beta \rho} Y_4^{\dagger \alpha \beta } Y_4^{\rho \sigma} \nonumber \\[1ex]
  &= \frac{3 M_{\phi_3^{1/3}}}{16 \pi} {\rm Tr} [ Y_4^\dagger Y_4 ] ,
\label{decaywidth}
\end{align}
due to the unitarity of mixing the matrices, the final result turns out to be independent of any mixing parameter when the fermion masses are neglected. The same simplification occurs for the decays of the vector leptoquark $X_\mu$ after summing over the family index.

\section{Heavy Gauge Boson Masses}
%
%
The VEV of the scalar $\chi$ corresponding to  $\langle \chi \rangle = \textrm{diag} ( 0, \, 0, \, 0, \, v_{\chi}/\sqrt{2})$ is responsible for the spontaneous breaking of $\SU(4)_C \otimes \SU(2)_L \otimes \U(1)_R \to \SU(3)_C \otimes \SU(2)_L \otimes \U(1)_Y$. After this spontaneous breaking, the gauge boson corresponding to the $\SU(4)_C$ generator $T_4$ ($B_\mu^{'}$) mixes with the $\U(1)_R$ gauge boson ($Z^{\mu}_R$), which results in the massless $\U(1)_Y$ gauge boson ($B_\mu$) of the Standard Model and an orthogonal massive state ($Z^{'}_\mu$) defined by
\begin{eqnarray}
\begin{pmatrix} Z_{R\mu} \\ B^{'}_{\mu} \end{pmatrix} = \begin{pmatrix} \cos \theta_4 & \sin \theta_4 \\ -\sin \theta_4 & \cos \theta_4 \end{pmatrix}
\begin{pmatrix} Z^{'}_{\mu} \\ B_{\mu} \end{pmatrix},
\end{eqnarray}
where the mixing angle and the hypercharge gauge coupling are given by
\begin{align}
	\sin{\theta_4} = \frac{g_4}{\sqrt{g_4^2 + \frac{2}{3}g_R^2}} \hspace{0.5cm} {\rm and} \hspace{0.5cm} g_Y = \frac{g_R g_4}{\sqrt{g_4^2 + \frac{2}{3}g_R^2}}.
\end{align}
The mass of the heavy neutral gauge boson reads as
\begin{align}
	M_{Z^{'}}^2 \simeq \frac{1}{4} \left(g_R^2 + \frac{3}{2}g_4^2\right) v_{\chi}^2,
\end{align}
while the mass of the vector leptoquark is given by
\begin{equation}
	M_X^2 \simeq \frac{g_4^2}{4} v_\chi^2.
\end{equation}
%
%
\section{Scalar Potential and Leptoquark Masses}
%
The scalar potential may be written as 
\begin{eqnarray}
V &\supset& \mu^2_{H_1} H_1^{\dagger} {H_1} + \mu_{\chi}^2 \chi^{\dagger} \chi + \mu_{\Phi}^2 {\rm Tr}\left[ \Phi^{\dagger} \Phi \right] + \lambda_1 H_1^{\dagger} {H_1} \chi^\dagger \chi + \lambda_2 H_1^{\dagger} H_1 {\rm Tr} \left[ \Phi^{\dagger} \Phi \right] + \lambda_3 \chi^\dagger \chi {\rm Tr} \left[ \Phi^{\dagger} \Phi \right] \\
&+& \left( \lambda_4 H_1^{\dagger} \chi^{\dagger} \Phi \chi + \mbox{h.c.}\right) + \lambda_5 H_1^{\dagger} {\rm Tr} \left[ \Phi \, \Phi^{\dagger} \right] H_1 + \lambda_6 \chi^{\dagger} \Phi \Phi^{\dagger} \chi + \lambda_7 \left( H_1^{\dagger} H_1\right)^2 + \lambda_8 \left( \chi^{\dagger} \chi \right)^2 \nonumber\\
&+& \lambda_9 {\rm Tr} \left[ ( \Phi^{\dagger} \Phi )^2 \right] + \lambda_{10} \left( {\rm Tr} [\Phi^{\dagger} \Phi] \right)^2 + \left( \lambda_{11} H_1^{a\dagger} {\rm Tr} [\Phi^a \Phi^b] H_1^{b\dagger} + \lambda_{12} H_1^{a\dagger} {\rm Tr} \left[ \Phi^a \Phi^b \Phi^{b\dagger}\right] \right. \nonumber\\ 
&+& \left. \lambda_{13} H_1^{a\dagger}{\rm Tr} \left[ \Phi^a \Phi^{b\dagger} \Phi^b \right] + \mbox{h.c.}\right) + \lambda_{14} \chi^{\dagger} \Phi^{\dagger} \Phi \chi + \lambda_{15} {\rm Tr} \left[ \Phi^{a\dagger} \Phi^b \Phi^{b\dagger} \Phi^a \right] + \lambda_{16} {\rm Tr} \left[ \Phi^{a\dagger} \Phi^b \right]{\rm Tr} \left[ \Phi^{b\dagger} \Phi^a \right] \nonumber \\[1ex] 
&+&  \lambda_{17}{\rm Tr}\left[ \Phi^{a \dagger} \Phi^{b \dagger}\right] {\rm Tr} \left[ \Phi^a \Phi^b \right] + \lambda_{18} {\rm Tr} \left[ \Phi^{a\dagger} \Phi^{b\dagger} \Phi^a \Phi^b \right] + \lambda_{19} {\rm Tr} \left[ \Phi^{a \dagger} \Phi^{b \dagger} \Phi^{b} \Phi^{a} \right] \, , \nonumber
\end{eqnarray}
\vspace{0.1cm}
where the trace is in $\SU(4)_C$ space, and $a, b$ are $\SU(2)_L$ indices. Taking the limit $v_\chi \gg v_1, v_2$ we find that
\begin{eqnarray}
M_{\Phi_8}^2 &=& \left( \frac{\sqrt{3} \lambda_4}{4}\cot{\beta} - \frac{3}{8}(\lambda_6 + \lambda_{14}) \right) v_{\chi}^2, \\
M_{\Phi_3}^2 &=& \left( \frac{\sqrt{3} \lambda_4}{4}\cot{\beta} + \frac{\lambda_{14} - 3\lambda_6}{8}\right) v_{\chi}^2, \\
M_{\Phi_4}^2 &=& \left( \frac{\sqrt{3} \lambda_4}{4}\cot{\beta} + \frac{\lambda_6 - 3 \lambda_{14}}{8} \right) v_{\chi}^2, \\
M_{H_2}^2 &=& \frac{\sqrt{3} \lambda_4}{4} \cot{\beta} \, v_{\chi}^2.
\end{eqnarray}
These mass expressions imply the following tree-level sum rule~\cite{Faber:2018qon}:
\begin{equation}
M_{\Phi_8}^2 + 2 M_{H_2}^2 = \frac{3}{2} \left( M_{\Phi_{3}}^2 + M_{\Phi_{4}}^2\right),
\end{equation}
which implies that the masses of the scalars in the theory are related and of the same scale.
%
%
\section{Leptoquark Feynman Rules}
\label{leptoquarks-rules}
%
The vector leptoquark has the following interactions
\beq
-\mathcal{L} \supset \frac{g_4}{\sqrt{2}} X_\mu (\bar{Q}_L \gamma^\mu \ell_L + \bar{u}_R \gamma^\mu \nu_R + \bar{d}_R \gamma^\mu e_R) + {\rm h.c.}\, .
\eeq
For the scalar leptoquarks, after expanding the $\SU(2)_L$ indices of the Yukawa interactions in Eq.~\eqref{eq:YukInt}, we can describe the interactions by
\begin{align}
-\mathcal{L} \supset & \,\, Y_4^{ji} \left[ \bar{d}_R ^i \left(\phi_3^{1/3}\right)^* \nu_L^j  + \bar{d}_R^i \left(\phi_3^{-2/3}\right)^* e_L^j  + \bar{e}_R^i \left(\phi_4^{5/3}\right)^* u_L^j  + \bar{e}_R^i \left(\phi_4^{2/3}\right)^* d_L^j\right] \nonumber \\[1.5ex]
& + Y_2^{ji} \left[\bar{N}^i \, \phi_3^{-2/3} \, u_L^j - \bar{N}^i \, \phi_3^{1/3} \, d_L^j + \bar{u}_R^i \, \phi_4^{2/3} \, \nu_L^j  - \bar{u}_R^i \, \phi_4^{5/3} \, e_L^j    \right],
\end{align}
where $i,j$ correspond to family indices. The mass eigenstates $\phi_A^{-2/3}$ and $\phi_B^{-2/3}$ are defined as
\begin{eqnarray}
    \phi_3^{-2/3}  &= & \cos \theta_\text{LQ} \, \phi_A^{-2/3} + \sin \theta_\text{LQ} \, \phi_B^{-2/3},  \\[1ex]
   (\phi_4^{2/3})^* &= & - \sin \theta_\text{LQ} \, \phi_A^{-2/3} + \cos \theta_\text{LQ} \, \phi_B^{-2/3}.
\end{eqnarray}
The masses of the leptoquarks have to be above 1 TeV from LHC constraints and the mixing is determined by the electroweak scale; therefore, the mixing angle $\theta_\text{LQ}$ is very small in general.
\\

The Feynman rules for the interactions of the leptoquarks with fermions are as follows:
\begin{itemize}
\item $X_\mu$:
\begin{eqnarray}
&&\bar{d}^i e^j X_\mu: \hspace{0.5cm} i \frac{g_4}{\sqrt{2}}  \left[  (V_{DE})^{ij} P_R  + (V_2)^{ji} P_L  \right] \gamma^\mu,  \\[1.5ex]
&&\bar{u}^i \nu^j X_\mu : \hspace{0.5cm}  i \frac{g_4}{\sqrt{2}}\left(K_1 V_{\rm CKM} K_2 V_{DE} K_3 V_{\rm PMNS} \right)^{ij} \gamma^\mu P_L, \\
&& \bar{u}^i N^j X_\mu : \hspace{0.5cm} i \frac{g_4}{\sqrt{2}} V_1^{ji} \gamma^\mu P_R.
\end{eqnarray}
\item $\phi_3^{1/3}$:
\begin{eqnarray}
&&\bar{\nu}^i d^j \phi_3^{1/3}: \hspace{0.5cm} i (V^*_4)^{ij} \, P_R,  \\[1.5ex]
&&\bar{N}^i d^j \phi_3^{1/3}: \hspace{0.5cm}  -i (V_3^{T} \, K_1 \, V_{\rm CKM} \, K_2)^{ij} \, P_L. \hspace{5cm}
\end{eqnarray}
\item $\phi_4^{5/3}$:
\begin{eqnarray}
&&\bar{u}^i e^j \phi_4^{5/3}: \hspace{0.5cm} i \left[ - (V_5^{T} \, V^{\dagger}_{\rm PMNS} \, K_3^*)^{ij} \, P_L + (V_6^*)^{ij} \, P_R\right]. \hspace{4.25cm}
\end{eqnarray}
\item $\phi_A^{-2/3}$:
\begin{eqnarray}
&&\bar{N}^i u^j \phi_A^{-2/3}: \hspace{0.5cm} i \cos{\theta_{\rm LQ}} (V_3^{T})^{ij} \, P_L, \\[1.5ex]
&&\bar{\nu}^i u^j \phi_A^{-2/3}: \hspace{0.5cm} - i \sin{\theta_{\rm LQ}} (V_5^{*})^{ij} \, P_R, \\[1.5ex]
&&\bar{e}^i d^j \phi_A^{-2/3}: \hspace{0.5cm} i \left( \cos{\theta_{\rm LQ}} (K_3 V_{\rm PMNS} V_4^{*})^{ij} P_R - \sin{\theta_{\rm LQ}} (V_6^{T} K_1 V_{\rm CKM} K_2)^{ij} P_L\right).
\end{eqnarray}
\item $\phi_B^{-2/3}$:
\begin{eqnarray}
&& \bar{N}^i u^j \phi_B^{-2/3}: \hspace{0.5cm} i \sin{\theta_{\rm LQ}} (V_3)^{ji} \, P_L, \\[1.5ex]
&& \bar{\nu}^i u^j \phi_B^{-2/3}: \hspace{0.5cm} i \cos{\theta_{\rm LQ}} (V_5^{*})^{ij} P_R, \\[1.5ex]
&& \bar{e}^i d^j \phi_B^{-2/3}: \hspace{0.5cm} i \left( \sin{\theta_{\rm LQ}} (K_3 V_{\rm PMNS} V_4^{*})^{ij} P_R + \cos{\theta_{\rm LQ}} (V_6^{T} K_1 V_{\rm CKM} K_2)^{ij} P_L\right),
\end{eqnarray}
\end{itemize}
where $P_{L,R}$ are the chiral projection operators $P_{L,R} = (1 \mp \gamma^5)/2$ and the interaction matrices are given by
\begin{align}
	V_1 & = N^{\dagger}_c U_c,  \hspace{2.1cm} V_2 = E^{\dagger}_c D_c, \hspace{1.7cm} V_3 = U^{T} Y_2 N_c,\\[1ex] 
V_4 & = N^T Y_4 D_c,  \hspace{1.6cm} V_5 = N^{T}Y_2 U_c,\hspace{1.4cm} V_6 = U^{T} Y_4 E_c, \nonumber \\[1ex] 
	U^{\dagger}D & = K_1 V_{\rm CKM} K_2,\hspace{0.7cm} E^{\dagger} N = K_3 V_{\rm PMNS}, \hspace{1.cm} V_{DE} = D^{\dagger} E, \nonumber
\end{align}
$K_1$ and $K_3$ are diagonal matrices containing three phases, $K_2$ is a diagonal matrix with two phases.
%
%
\section{Higgs Feynman Rules}
\label{sec:appHiggs}
The Feynman for the physical scalars in the two Higgs doublets rules correspond to:
\begin{itemize}
\item $H^+$:
\begin{eqnarray}
&&\bar{u}^i d^j H^+: \hspace{0.5cm} i \left[ (C_{Lud})^{ij} P_L + (C_{Rud})^{ij} P_R \right],  \\[1.5ex]
&&\bar{N}^i e^j H^+ : \hspace{0.5cm}  i ( C_{Ne} )^{ij} P_L ,\\[1.5ex]
&& \bar{\nu}^i e^j H^+ : \hspace{0.5cm} - i ( C_{\nu e} )^{ij} P_R .
\end{eqnarray}
\item $H^-$:
\begin{eqnarray}
&& \bar{d}^i u^j H^-: \hspace{0.5cm} i \left[ (C_{Rud}^*)^{ji} P_L + (C_{Lud}^*)^{ji} P_R \right],  \\[1.5ex]
&&\bar{e}^i N^j H^-: \hspace{0.5cm}  i (C_{Ne}^*)^{ji} P_R, \\[1.5ex]
&&\bar{e}^i \nu^j H^-: \hspace{0.5cm}  -i (C_{\nu e}^*)^{ji} P_L. 
\end{eqnarray}
\item $h$:
\begin{eqnarray}
&& \bar{u}^i u^j h: \hspace{0.5cm}  i \left[ (C_{uu}^h)^{ij} P_L + (C_{uu}^{h*})^{ji} P_R \right],\\[1.5ex] 
&& \bar{N}^i \nu^j h: \hspace{0.5cm}  i \left[ (C_{N\nu}^h)^{ij} P_L + (C_{N\nu}^{h*})^{ji} P_R \right],\\[1.5ex] 
&& \bar{d}^i d^j h: \hspace{0.5cm}  i \left[ (C_{dd}^h)^{ij} P_L + (C_{dd}^{h*})^{ji} P_R \right],\\[1.5ex] 
&& \bar{e}^i e^j h: \hspace{0.5cm}  i \left[ (C_{ee}^h)^{ij} P_L + (C_{ee}^{h*})^{ji} P_R \right] .
\end{eqnarray}
\item $H$:
\begin{eqnarray}
&& \bar{u}^i u^j H: \hspace{0.5cm}  i \left[ (C_{uu}^H)^{ij} P_L + (C_{uu}^{H*})^{ji} P_R \right],\\[1.5ex] 
&& \bar{N}^i \nu^j H: \hspace{0.5cm}  i \left[ (C_{N\nu}^H)^{ij} P_L + (C_{N\nu}^{H*})^{ji} P_R\right],\\[1.5ex] 
&& \bar{d}^i d^j H: \hspace{0.5cm}  i \left[ (C_{dd}^H)^{ij} P_L + (C_{dd}^{H*})^{ji} P_R \right],\\[1.5ex] 
&& \bar{e}^i e^j H: \hspace{0.5cm}  i \left[ (C_{ee}^H)^{ij} P_L + (C_{ee}^{H*})^{ji} P_R \right] .
\end{eqnarray}
\item $A$:
\begin{eqnarray}
&& \bar{u}^i u^j A: \hspace{0.5cm}   (C_{uu}^A)^{ij} P_L - (C_{uu}^{A*})^{ji} P_R\,, \\[1.5ex]
&& \bar{N}^i \nu^j A: \hspace{0.5cm}   (C_{N\nu}^A)^{ij} P_L - (C_{N\nu}^{A*})^{ji} P_R\, , \\[1.5ex]
&& \bar{d}^i d^j A: \hspace{0.5cm}   (C_{dd}^A)^{ij} P_L - (C_{dd}^{A*})^{ji} P_R\, , \\[1.5ex]
&& \bar{e}^i e^j A: \hspace{0.5cm}   (C_{ee}^A)^{ij} P_L - (C_{ee}^{A*})^{ji} P_R\, .
\end{eqnarray}
\end{itemize}
where the interaction matrices are given by
\begin{align}
C_{Lud} & = U_c^T \left( Y_1^T \sin \beta  - Y_2^T \frac{\cos \beta}{2 \sqrt{3}} \right) D, \hspace{1.1cm}
C_{Rud} = - U^\dagger \left( Y_3^* \sin \beta -Y_4^*  \frac{\cos \beta}{2 \sqrt{3}} \right) D_c^* , \nonumber \\[1.5ex]
C_{Ne} & = N_c^T \left( Y_1^T \sin \beta + Y_2^T \frac{\sqrt{3} \cos \beta}{2} \right) E, \hspace{0.8cm}
C_{\nu e} = N^\dagger \left(  Y_3^* \sin \beta  + Y_4^* \frac{\sqrt{3} \cos \beta}{2} \right) E_c^*, \nonumber \\[1.5ex]
C_{uu}^H & = U^T_c \left( Y_1^T \frac{\cos \alpha}{\sqrt{2}} + Y_2^T \frac{\sin \alpha}{ 2\sqrt{6}} \right)   U , \hspace{1.3cm}
C_{N\nu}^H = N_c^T \left(  Y_1^T \frac{\cos \alpha}{\sqrt{2}} - Y_2^T \frac{3 \sin \alpha}{2\sqrt{6}} \right) N , \nonumber \\[1.5ex]
C_{dd}^H & = D_c^T \left( Y_3^T \frac{\cos \alpha}{\sqrt{2}} + Y_4^T \frac{\sin \alpha}{ 2 \sqrt{6}}  \right) D , \hspace{1.3cm}
C_{ee}^H = E_c^T \left( Y_3^T \frac{\cos \alpha}{\sqrt{2}} - Y_4^T \frac{3\sin \alpha}{2\sqrt{6}}  \right) E, \nonumber \\[1.5ex]
C_{uu}^h & = U^T_c \left( -Y_1^T \frac{\sin \alpha}{\sqrt{2}} + Y_2^T \frac{\cos \alpha}{ 2 \sqrt{6}} \right)   U , \hspace{0.95cm}
C_{N\nu}^h = N_c^T \left( -Y_1^T \frac{\sin \alpha}{\sqrt{2}} - Y_2^T \frac{3 \cos \alpha}{2\sqrt{6}} \right) N , \nonumber \\[1.5ex]
C_{dd}^h & = D_c^T \left( -Y_3^T \frac{\sin \alpha}{\sqrt{2}} + Y_4^T \frac{\cos \alpha}{ 2\sqrt{6}}  \right) D , \hspace{1cm}
C_{ee}^h = E_c^T \left( -Y_3^T \frac{\sin \alpha}{\sqrt{2}} - Y_4^T \frac{3\cos \alpha}{2\sqrt{6}}  \right) E, \nonumber \\[1.5ex]
C_{uu}^A & = U_c^T \left( Y_1^T \frac{\sin \beta}{\sqrt{2}} - Y_2^T \frac{\cos \beta}{2 \sqrt{6}}  \right) U ,\hspace{1cm}
C_{N\nu}^A = N_c^T \left( Y_1^T \frac{\sin\beta}{\sqrt{2}} + Y_2^T \frac{3\cos\beta}{2\sqrt{6}} \right)  N , \nonumber \\[1.5ex]
C_{dd}^A & = D_c^T \left( - Y_3^T \frac{\sin \beta}{\sqrt{2}} + Y_4^T \frac{\cos \beta}{  2 \sqrt{6}}  \right) D  , \hspace{1cm}
C_{ee}^A = E_c^T \left( - Y_3^T \frac{\sin \beta}{\sqrt{2}} - Y_4^T \frac{3 \cos \beta}{ 2 \sqrt{6}}  \right) E . \nonumber
\end{align}

\end{widetext}

\bibliography{LQ-relations}

\end{document}